\documentclass[trackchanges]{aastex63}
\usepackage[nohyperlinks, nolist]{acronym}
\usepackage{upgreek}
\usepackage{ulem}

\definecolor{sof}{rgb}{0.4, 0.0, 0.6}
\definecolor{mac}{rgb}{0.0, 0.5, 0.0}
\definecolor{pin}{rgb}{1.0, 0.5, 0.0}

\newcommand{\PSUAA}{Department of Astronomy and Astrophysics, The Pennsylvania State University, University Park, PA, 16802, USA}
\newcommand{\PSUCEHW}{Center for Exoplanets and Habitable Worlds, The Pennsylvania State University, University Park, PA, 16802, USA}
\newcommand{\PSETI}{Penn State Extraterrestrial Intelligence Center, The Pennsylvania State University, University Park, PA, 16802, USA}
\newcommand{\UWM}{Department of Astronomy, University of Wisconsin--Madison, Madison, WI, USA}
\newcommand{\COL}{Department of Religion, Columbia University, New York, NY, USA}
\newcommand{\GODDARD}{NASA Goddard Space Flight Center}
\newcommand{\ROCH}{Department of Physics and Astronomy, University of Rochester, University of Rochester}

\usepackage{hyperref}
\hypersetup{
    colorlinks=true,
    linkcolor=blue,
    filecolor=magenta,      
    urlcolor=cyan
    }

\shorttitle{Earth Detecting Earth}
\shortauthors{Sheikh et al.}

\begin{document}

\title{Earth Detecting Earth: At what distance could Earth's constellation of technosignatures be detected with present-day technology?}

\correspondingauthor{Sofia Z. Sheikh}
\email{ssheikh@seti.org}

\author[0000-0001-7057-4999]{Sofia Z. Sheikh}
\affiliation{SETI Institute, 339 N Bernardo Ave, Suite 200, Mountain View, CA 94043}
\affiliation{\PSETI}
\affiliation{Breakthrough Listen, University of California, Berkeley, 501 Campbell Hall 3411, Berkeley, CA 94720, USA}

\author[0000-0003-4591-3201]{Macy J. Huston}
\affiliation{\PSETI}
\affiliation{Astronomy Department, University of California, Berkeley, CA, 94720, USA}

\author[0000-0003-3988-9022]{Pinchen Fan}
\affiliation{\PSETI}
\affiliation{\PSUAA}
\affiliation{\PSUCEHW}

\author[0000-0001-6160-5888]{Jason T.\ Wright}
\affiliation{\PSETI}
\affiliation{\PSUAA}
\affiliation{\PSUCEHW}

\author[0000-0002-9539-4203]{Thomas Beatty}
\affiliation{\UWM}

\author[0000-0001-9790-7554]{Connor Martini}
\affiliation{\COL}

\author[0000-0002-5893-2471]{Ravi Kopparapu}
\affiliation{\GODDARD}

\author[0000-0002-4948-7820]{Adam Frank}
\affiliation{\ROCH}

\turnoffedit

\begin{abstract}
    The field of \ac{SETI} searches for ``technosignatures'' that could provide the first detection of life beyond Earth through the technology that an \ac{ETI} may have created. Any given \ac{SETI} survey\edit1{, if no technosignatures are detected, should set upper limits based on the kinds of technosignatures it should have been able to detect}; the sensitivity of many SETI searches requires that their target sources (e.g., Dyson spheres or Kardashev II/III level radio transmitters) emit with power far exceeding the kinds of technology humans have developed. In this paper, we instead turn our gaze Earthward, minimizing the axis of extrapolation by only considering transmission and detection methods commensurate with an Earth-2024 level. We evaluate the maximum distance of detectability for various present-day Earth technosignatures --- radio transmissions, atmospheric technosignatures, optical and infrared signatures, and objects in space or on planetary surfaces --- using only present-day Earth instruments, providing one of the first fully cross-wavelength comparisons of the growing toolbox of \ac{SETI} techniques. In this framework, we find that Earth's space-detectable signatures span 13 orders of magnitude in detectability, with intermittent, celestially-targeted radio transmission (i.e., planetary radar) beating out its nearest non-radio competitor by a factor of $10^3$ in detection distance. This work highlights the growing range of ways that exoplanet technosignatures may be expressed, the growing complexity and visibility of the human impact upon our planet, and the continued importance of the radio frequencies in \ac{SETI}.
\end{abstract}

\keywords{SETI --- 
technosignatures --- exoplanets}

\acresetall

\section{Introduction}
\label{sec:introduction}

\Ac{SETI} began with an important revelation: humans on Earth can produce electromagnetic signals that our own telescopes could detect at interstellar distances. When \citet{Cocconi1959} and \citet{Drake1961} first proposed and executed searches for artificial radio signals from nearby stars, a key motivation was that recently-built and under-construction radio telescopes had the requisite size and power to, in theory, communicate with one another at distances on the order of 10 ly. \citet{Schwartz1961} calculated just before the invention of the laser that the same could be done at optical wavelengths. Since then, a continuing theme in the discipline of \ac{SETI} is that the invention of new technologies on Earth influences how we search for signatures of \ac{ETI} beyond Earth.

For the purposes of SETI, \citet{Wright2018} proposed the following definition of the word {\it technology}: ``The physical manifestations of deliberate engineering. That which produces a technosignature.'' In this work, we will focus on technologies that produce technosignatures that may conceivably be observable across interstellar distances, regardless of whether we have the capabilities to observe them at this time. Therefore, we will begin with a brief history of space-detectable human technologies.

While the word ``technology'' evokes the present century, some millennia-old human technologies had deep and potentially remotely observable effects, which have already been recognized by scholars in the Anthropocene and ``Big History'' literature \citep[e.g., ][]{crutzen2000}. One potential candidate for the ``first remotely-detectable technosignature'' is the methane-recession that led to the Younger Dryas glacial climate episode 12.8~kya, potentially influenced by the human extinction of megafauna \citep{lewis2018human}. This was followed by the gradual development and parallel invention of agriculture across the continents, which may have had detectable impacts on the carbon dioxide and methane levels in the atmosphere starting around 6--7~kya \citep{lewis2015defining}. \edit1{The human link to these early signatures is somewhat tentative and still debated --- even with the significant planetary context we have for Earth --- which complicates their interpretation as technosignatures. However,} the human involvement in the constellation of industry-induced changes known as the ``Great Acceleration'' \citep{crutzen2006anthropocene}, encompassing a wide range of measurable atmospheric, biogeochemical, and land use changes \citep[e.g., ][]{Haqq-Misra2022agriculture}, is unambiguous.

In contrast with these unintentional remote technosignatures, Table \ref{tab:timeline} shows an approximate timeline of the human development of communication technologies and telescopes. In 1895, Guglielmo Marconi made the first communicative radio transmission, carrying a Morse code signal \citep{Marconi1897, Bondyopadhyay1995}. Karl Jansky effectively created the first radio telescope in 1932 \citep{Jansky1932, Jansky1933}. While infrared radiation from the Sun and Moon had already been detected, the 1950s can be regarded as the beginning of modern infrared astronomy, made possible by lead sulfide detectors \citep{Walker2000}. In 1979, \citet{Gfeller1979} introduced the use of infrared wavelengths for wireless communication. While Galileo Galilei is credited with the first astronomical use of a telescope, \cite{zuidervaart2000} described a complicated and controversial history of the device's invention in the year 1608. The photophone, invented by Alexander Graham Bell and Charles Sumner Tainter in 1880, was the first application of optical transmissions for communication \citep{Bell1880}, beyond fire or static mirrors\footnote{Because humans can detect optical wavelengths without technological assistance, the history of transmission and reception in this wavelength is more complicated, and arguments could be made for earlier technologies e.g., the heliotrope or heliograph.}.

Higher frequencies of light cannot travel from space through Earth's atmosphere, so detecting ultraviolet, X-ray, and gamma-ray radiation from space requires rocket launches. German V-2 rockets launched from the Naval Research Laboratory first detected solar UV radiation in 1946 \citep{Durand1947} and solar X-ray radiation in 1949 \citep{Friedman1951}. While these early sounding rocket experiments set upper limits on solar gamma rays, it was not until 1961 that gamma rays were first detected in space with the Satellite Explorer XI \citep{Kraushaar1962}. The Naval Research Laboratory first presented UV communications in the 1930s \citep{Guo2021}. The use of X-rays for communication was first proposed by Keith Gendreau in 2007 \citep{li2022}, and NASA performed a test of such technology in space in 2019\footnote{\url{https://techport.nasa.gov/view/94821}}. Gamma-ray communications were only proposed very recently by \citet{li2022}. 

In addition to electromagnetic technologies, we also consider multi-messenger approaches. The first neutrino detector for neutrinos from space (in this case, the Sun) was created in 1965 by Ray Davis, using an underground tank in the Homestake Gold Mine \citep{DAVIS1994}. Although humans do not commonly use neutrinos for communication, \citet{Stancil2012} demonstrated the technological capability in 2012. The search for gravitational waves began with the Laser Interferometer Gravitational-Wave Observatory (LIGO) in 1999\footnote{\url{https://www.ligo.caltech.edu/page/facts}}, with the update to Advanced LIGO completed in 2014, allowing for the first detection in 2015 \citep{gw150914}. Humans have not yet produced communicative gravitational waves.

\begin{table}[ht]
    \centering
    \begin{tabular}{|c|c|c|}
        \hline
        Signal Type & Humans Create Telescopes/Detectors & Humans Create Transmitters \\
        \hline
        Radio & 1932 & 1895  \\ 
        Infrared & 1950s & 1979 \\ 
        Optical & 1609 & 1880 \\ 
        Ultraviolet & 1946 & 1930s \\ 
        X-ray & 1949 & 2019 \\ 
        Gamma ray & 1961 & -  \\ 
        Neutrino & 1967 & 2012\\ 
        Gravitational wave & 2014 & -  \\ 
        \hline
    \end{tabular}
    \caption{This table traces a rough timeline through the development of human technologies across wavelength ranges and multi-messenger approaches. While the exact ``firsts'' of certain technologies can be difficult to define, we attempt here to roughly outline two key milestones. The middle column shows when humans first created a transmitter for communication of information via a given signal type. The leftmost column shows when humans first created telescopes or detectors capable of detecting each signal type from space.}
    \label{tab:timeline}
\end{table}

One takeaway from Table \ref{tab:timeline} is that the development of transmitters and receivers tends to be (relatively) temporally linked across wavelengths and messengers. For some technologies, we see that once humans figure out how to generate a particular signal, applying the reception technology for that signal to space is not far off. On the other hand, some technologies that began first with telescopes are now being developed in our efforts to communicate effectively with our own space technology throughout the Solar System. 

Using Earth as a mirror in this way, we can employ the concept of the ichnoscale ($\iota$) from \citet{Socas-Navarro2021} --- ``the relative size scale of a given technosignature in units of the same technosignature produced by current Earth technology.'' An $\iota$ value of 1 is defined by Earth's current technology. This necessarily evolves over time --- for this work, we set the ichnoscale to Earth-2024 level technology, including near-future technologies that are already in development.

This is equivalent to pinning the extrapolation axis of the Nine Axes of Merit for SETI searches \citep{sheikh2020merits} to its most favorable value: evaluating all human technosignatures that could be observable from space (even at very short, intra-solar system distances) with the strong constraint that these signatures can be no stronger than they have been at some point in recorded Earth history. One could imagine a spaceship loaded with all of humanity's best modern-day instruments, launched towards a mirror image of Earth --- which present-day technosignatures would be detectable first? Which impacts of our activities would be detectable at interstellar distances, and for which would the ship need to wait until it entered the mirror system, or even resolved the mirror Earth itself? 

\edit1{It should be noted that the idea of a ``mirror Earth'' or Earth-level \ac{ETI} not meant to be taken literally but is instead a helpful framing device --- there is no universal technological level or single civilization on Earth, we cannot assume that convergent technological or biological evolution would be at play across the galaxy, and the ``present'' level of technology rapidly changes.}

In this work, we use Earth's current technological capabilities to establish quantitative standards for technosignature detectability. For the Earth-detecting-Earth approach, we apply the ichnoscale value of 1 for both transmitters and detectors ($\iota_{\rm TS} = \iota_{\rm detector} =1$) and calculate the distance $d_{\rm detect}$ where a definitive detection (which we define as a signal-to-noise ratio of 5) could be made for an array of technosignatures. This should not be taken as a comprehensive list of every technosignature, as it does not account for classes of signatures we haven't yet managed to generate on Earth (e.g., megastructures or antimatter propulsion signatures). 

In Section \ref{sec:radio_transmissions}, we examine the detectability of radio transmissions. In Section \ref{sec:atmospheric}, we discuss atmospheric technosignatures.  In Section \ref{sec:optical}, we assess technosignatures in the optical and infrared. In Section \ref{sec:objects}, we explore the detectability of objects in space or on other planetary surfaces. In Section \ref{sec:discussion}, we discuss the implications of these calculations in the search for technosignatures and some technosignatures we produce but could not currently detect. Finally, we conclude and provide an illustration of the Earth-detecting-Earth spatial limits in Section \ref{sec:conclusion}.

\section{Radio Transmissions}
\label{sec:radio_transmissions}

Humans use radio frequencies for communication in order to efficiently transmit information at lightspeed through common atmospheric phenomena on Earth (i.e., clouds) over long distances. Radio transmitters are often directed, or beamed, to increase power in a particular direction; however, if we normalize by scaling to the \ac{EIRP}, commonly encountered transmitters in everyday life span several orders of magnitude.

In this section, we calculate the detectability of four classes of radio emission from human technosignatures:

\begin{enumerate}
    \item \textbf{Intermittent, celestially-targeted radio transmission}: E.g., intentional beamed \ac{METI} signals like the Arecibo Message \citep{arecibomessage1974},  planetary-scale radar transmissions for asteroid and planetary characterization (discussed further in Section \ref{sec:objects}). Arecibo's characteristic \ac{EIRP} at S-band is 20~TW \citep{ekers2002seti}.
    
    \item \textbf{Persistent, celestially-targeted radio transmission}: E.g., NASA's Deep Space Network (DSN). DSN is used to frequently communicate between space probes (e.g. Martian orbiters, space telescopes, etc.) and ground stations. The ground station uplink is usually at 20 kW in S/X-bands with 70-meter radio dishes, although it can transmit at 400 kW in the S-band \citep{dsn_tech}. This translates to an EIRP of 965~MW using the 400 kW transmitter in the S-band.
    
    \item \textbf{Persistent, omnidirectional radio leakage}: E.g., cell towers, television broadcasters, radio stations. These transmitters are fixed on the surface of Earth and are not evenly distributed across the surface of the planet \citep{sullivan1978eavesdropping}, thereby revealing the rotation period of the Earth and the distribution of transmitters across its surface. In addition, integrating each individual transmitter's power, for a particular technology in a particular allocated band, will result in a total power visible at some distance from Earth. We use a value of 4~GW from \citet{saide2023simulation}, which is the peak detected power of Earth's LTE mobile technology at a favorable vantage\footnote{This value is likely an underestimate, given that it is not intended to be interpreted as an \ac{EIRP}, but converting to a true EIRP would require a significant independent investigation, taking into about beaming angles, terrain, the contributions from towers versus handsets, and other complex considerations in an integrated incoherent sum that is outside the scope of this paper}.
    
    \item \textbf{Radio signals from artifacts}: E.g., downlinks from planetary orbiters or other space assets. These transmitters are intended to efficiently communicate with radio observatories on Earth's surface across near-interstellar distances, and these engineering constraints provide an insightful proxy for the potential design of transmitters on an \ac{ETI} probe in the solar system. Here we use a characteristic \ac{EIRP} of 1.32~MW, which is the strength of the transmitter on the Voyager spacecraft \citep{posner1990voyager}.
\end{enumerate}

All of these classes of transmitters use the following transmission/detection physics.\footnote{Pulsed, broadband transmitters such as aircraft surveillance radar and military radar could also be considered here, but a) their detection physics are more complex due to the pulsed modulation and b) they fall between Case 2 and Case 3 in \ac{EIRP}, and thus will not represent a larger detection distance than the cases already considered.} The radiometer equation from \citet{enriquez2017breakthrough} gives the following spectral flux density $S_{min}$ for a narrowband source:

\begin{equation}
    \label{eq:radiometer_narrow}
    S_{min,narrow} = (\mathrm{SNR})\frac{\mathrm{SEFD}}{\Delta \nu_t}\sqrt{\frac{\Delta \nu}{n_{pol}\tau_{obs}}}
\end{equation}

\noindent where SNR is the signal-to-noise ratio, SEFD is the system equivalent flux density in e.g., Jy, $\Delta \nu_t$ is the transmitter bandwidth, $\Delta \nu$ is the frequency channelization of the data, $n_{pol}$ is the number of polarizations, and $\tau_{obs}$ is the observing time. The narrowband assumption is correct for many of these cases. The Arecibo message covered $\sim$10~Hz in bandwidth, but the finest frequency resolution used for radar science targets is about 0.01~Hz, which is what we will use here for Cases 1 and 2. For Case 4, looking at Voyager specifically, we can assume a 1~Hz wide signal. However, radio leakage from LTE mobile technology (Case 3) is wider-band, at up to 20~MHz \citep{saide2023simulation}. If we assume that we can match the channelization to the transmitted signal bandwidth ($\Delta \nu_t = \Delta \nu$) in the narrowband case, then the cases converge. We can convert between a spectral flux density and an \ac{EIRP} using:

\begin{equation}
    \label{eq:eirp}
    \mathrm{EIRP}_{min} = 4 \pi d^2 S_{min} \Delta\nu
\end{equation}

where $\mathrm{EIRP}_{min}$ is a normalized transmitter power metric indicating the weakest transmitter the search would be sensitive to, and $d$ is the distance between the transmitter and the receiver. If we then additionally assume a single polarization, we can combine Equations \ref{eq:radiometer_narrow} and \ref{eq:eirp} to:

\begin{equation}
    \label{eq:final_radio_distance_narrow}
    d = \sqrt{\frac{\mathrm{EIRP}_{min}}{4\pi (\mathrm{SNR})(\mathrm{SEFD})} \sqrt{\frac{\tau_{obs}}{\Delta\nu}}}
\end{equation}

The best near-term radio receiver on Earth is the \ac{SKA1-Mid} \citep{braun2017anticipated}, whose sensitivity can be calculated using the combined effective area over the system temperature. Using the 133 \ac{SKA1-Mid} 15m dishes and the 64 MeerKAT 13.5m dishes, and including S-band (which will optimistically be part of the development plan; to account for the specific frequency-band of Case 1), we find an \ac{SEFD} of 1.5~Jy from the system temperature and effective area given in Table~4 of \citet{braun2017anticipated}. The target \ac{SNR} of 5 is commensurate with the most sensitive narrowband technosignature searches \citep[e.g., ][]{sheikh2023green}. 

The final parameter is $\tau_{obs}$, or the total integration time. Standard observation times in radio technosignature surveys vary, but a current standard for single-dish observations is 3 5-minute integrations per target \citep[as used by the \ac{BL} group, e.g., ][]{franz2022breakthrough}. For a realistic extrapolation, an on-target time of 1 hour is feasible for ground-based observatories for most of their visible targets and is less than the period of most expected exoplanet rotations and orbits \citep{sheikh2019choosing}. 

It should be noted that the calculations in this section neglect Doppler drift rate and spectral broadening considerations; the former impacts extremely long integrations, while the latter could be significant for the narrowband signals past a few kpc. For example, in an extreme case, \citet{mingo2022accretion} performed a 164~hr integration for the \ac{LOFAR} deep-field; we opt not to consider 100s of hour integrations in this section due to these additional technical constraints and the general rarity of $>100$ hour single-target campaigns.

With these constraints, we get the following maximum detectable distances, $d_{\rm detect, radio}$, for the four cases:
\begin{enumerate}
    \item Case 1 (planetary radar): 12000~ly
    \item Case 2 (DSN): 65~ly
    \item Case 3 (LTE): 4.0~ly
    \item Case 4 (Voyager): 0.97~ly
\end{enumerate}

Note that the distances for Cases 1 and 2 in lightyears exceed the amount of time that these systems have been transmitting in years --- thus, while these signals \textit{will} be detectable at these distances in the future with this configuration, the signals from the systems have not yet reached those distances.

\section{Atmospheric Technosignatures}
\label{sec:atmospheric}

In an era of accelerating impacts from anthropogenic climate change, humanity's technological contributions to Earth's atmospheric composition, especially with regards to carbon dioxide (CO$_2$), have garnered significant public and policy interest \citep{ipcc2023climate}. However, while CO$_2$ has many sources, some technological, some biotic, and some abiotic \citep[which has led to it being investigated as a potential biosignature, or, more recently, as a biosignature pair with methane as in][]{schwieterman2018exoplanet}, there are other atmospheric technosignatures in Earth's atmospheres that have very few, or even no known non-technological sources. For example, \ac{CFCs}, a sub-category of halocarbons, are directly produced by human technology (with only very small natural sources), e.g., refrigerants and cleaning agents, and their presence in Earth's atmosphere constitutes a nearly unambiguous technosignature \citep{haqq2022cfc}. Nitrogen dioxide (NO$_2$), like CO$_2$, has abiotic, biogenic, and technological sources, but combustion in vehicles and fossil-fueled power plants is a significant contributor to the NO$_2$ in Earth's atmosphere \citep{kopparapu2021nitrogen}. The feasibility of detecting these two technosignature gases in the atmospheres of exoplanets has been explored by \citet{kopparapu2021nitrogen} and \citet{haqq2022cfc}, but we must adjust some of the model parameters for this paper's application of an $\iota_{\rm TS} = 1$ Earth-like system.

The best near-term ($\sim$ decades) technology for getting a spectrum of an Earth-like exoplanet's atmosphere will be direct-imaging coronagraphy via an upcoming 6m-class infrared/optical/ultraviolet telescope described in the 2020 Astronomy \& Astrophysics Decadal Survey \citep{national2021pathways}, which combines aspects of the LUVOIR and HabEx mission concepts \citep{luvoir2019, habex2020}.\footnote{Transmission spectroscopy is also widely used for close-in exoplanets, but is unlikely for our mirror Earth, which orbits a mirror sun every Earth year.} The currently planned mission is now called the Habitable Worlds Observatory (HWO). We will focus this case study on an Earth-like planet around a Sun-like star --- often, estimates of spectroscopic detectability are calculated assuming M-dwarf hosts, to optimize the ratio between the star and the planet's spectral contributions \citep[e.g., ][]{haqq2022cfc}. Here, we will use NO$_2$, an atmospheric Earth technosignature whose spectral signature lies within HWO's expected wavelength range, with spectral features from 0.2--0.7$\upmu$m \citep{kopparapu2021nitrogen}. According to the United States Environmental Protection Agency, the highest concentration of global NO$_2$ occurred in 1980, with a concentration of 113 ppb\footnote{The global concentration is 2--3 times lower at present than at the peak, see \url{https://www.epa.gov/air-trends/nitrogen-dioxide-trends}. \edit1{In general, in this work, we select the most extreme detection technologies and observed technosignatures that have been present on Earth, even if they are no longer extant.}}. As in the previous section, we continue to use our standard of \ac{SNR} = 5 for a detection. Following \citet{beatty2022detectability}, we adopt 300 hours as a standard observation time for an HWO-like mission target.

Using these parameters, we can scale the results from \citet{kopparapu2021nitrogen}, which used a 1D photochemical model and a synthetic spectrum generator \citep[the Planetary Spectrum Generator, ][]{villanueva2018planetary} to assess NO$_2$’s detectability as an atmospheric technosignature. Keeping the assumption of a cloud-free atmosphere, we find a maximum detectable distance of $d_{\rm detect, atmosphere}$ = 5.71~lyr. We discuss the possibility of time monitoring of changes in atmospheric technosignatures in Section~\ref{ssec:atmosphere_discussion}.

\section{Optical and Infrared Emission}
\label{sec:optical}

\subsection{City Lights}
\label{ssec:city_lights}
One of the strongest spectroscopic technosignatures present on Earth’s nightside is the emission from city lights. Conceptually, city lights are a compelling technosignature. The spectroscopic signature of Earth's city lights, as generated by high-pressure sodium lamps, is unique and difficult to cause via natural processes on a habitable terrestrial planet. The possible presence of city lights requires no extrapolation from current conditions on Earth, and they should be relatively long-lived for an urbanized civilization. This places the emission from nightside city lights high on the list of potential technosignatures to consider \citep{sheikh2020merits}. 

Indeed, the dominant light source on Earth's nightside is emission from street lights (or other area illumination lights), which reflect off nearby concrete and asphalt, as seen through the atmosphere from above. Modern street lights generally use high-pressure sodium lamps of several hundred Watts. As observed from space, the emission spectrum from city lights on Earth is predominantly this sodium emission spectrum convolved with the reflection spectrum of the nearby ground (typically concrete and asphalt on Earth) and filtered through the atmosphere's transmission function. The observable signature of night side city lights on a nearby exoplanet will therefore depend upon the type of lamp used for the lights, the reflection spectrum of the ground nearby, the transmission of the planetary atmosphere, and the overall urbanization level of the exoplanet.

The idea of searching for city lights on exoplanets was first suggested over a decade ago \citep{schneider2010far, loeb2012lights} and the detectability of sodium emission from city lights was recently considered in more detail in \citet{beatty2022detectability}. This recent analysis indicated that city lights could be feasibly detected in direct imaging searches for biosignatures on nearby exoplanets using next-generation space telescopes such as LUVOIR and HabEx. It was discovered that, while Earth-like urbanization fractions would not be detectable by any of the LUVOIR or HabEx architectures, the detection of \edit1{larger urbanization fractions, i.e., city light illumination several times that of Earth,} would be possible.

Using the formalism of \cite{beatty2022detectability}, we estimate that city lights on an exoplanet with an Earth-like urbanization fraction would not be detectable outside the Solar System by a 6m LUVEx observatory. On present-day Earth, 0.05\% of the planet's surface area is covered by strongly emitting urban centers. To detect the emission from this level of urbanization using 300 hours of time on LUVEx, an Earth-analog system would need to be at a distance of $d_{\rm detect, citylights}$ = 0.036 lyr for a detection at \ac{SNR} = 5. This is 2275 AU, which is close to the inner edge of the Oort Cloud. 

\subsection{Celestially-targeted Lasers}
\label{ssec:lasers}
The ``Fundamental Theorem of Optical SETI" is that Earth-2000 technology can generate laser pulses to outshine the Sun by orders of magnitude \citep{Horowitz2001}. However, humans do not regularly transmit such lasers into space. In contrast to radio transmissions, we find no record of any METI signals sent via laser.

On Earth, we produce a couple of optical and infrared lasers that are directed into space. First, ground-based telescope Adaptive Optics (AO) systems sometimes use Laser Guide Stars (LGSs) to correct for the atmosphere's effects on optical light from space.
Second, optical/infrared communication systems have been developed as an alternative to radio transmissions for transmitting large data volumes between the ground and space. We use these communication technologies for the following calculation, as they are more powerful than present AO lasers and have been developed explicitly to transmit through the atmosphere, rather than to be bounced off of an atmospheric layer.

Our transmitter is the Deep Space Optical Communications\footnote{\url{https://www.nasa.gov/mission/deep-space-optical-communications-dsoc/}} ground-based uplink laser array \citep[ULA, technical details from][]{Nau2022,Srinivasan2023}. The DSOC laser beam is centered at 1064 nm, with a spectral line width of $\lesssim0.25$ nm. It combines multiple lasers to achieve a total power of 7 kW with beam divergence as low as 40 microradians. 

Our receiver is the NIRSPEC instrument \citep{McLean1998} on Keck II. With specifications from the Keck/NIRSPEC Website\footnote{\url{https://www2.keck.hawaii.edu/inst/nirspec/}}, we adopt the low-resolution mode with r=2,500. We adopt a total integration time of 10 hours, on the order of a single night equivalent observing time. 

We consider a laser technosignature detectable at SNR=5 using the following equation:
\begin{equation}
    {\rm SNR} = \frac{n_{\rm laser}}{\sqrt{n_{\rm total}}} = \frac{A_{\rm tel} \epsilon t_{\rm obs} f_{\rm laser}}{\sqrt{A_{\rm tel} \epsilon t_{\rm obs}(f_{\rm laser} + f_{\rm background}) + r_{\rm dark} t_{obs} + n_{\rm read}}},
\end{equation}
where $n_{\rm laser}$, $n_{\rm total}$, and $n_{\rm read}$ are the numbers of detected photons from the laser, all sources, and read noise; $A_{\rm tel}$ is the collecting area of the telescope; $\epsilon$ is the quantum efficiency of the detector; $t_{\rm obs}$ is the total integration time; $f_{\rm laser}$ and $f_{\rm background}$ are photon fluxes from the laser and the background; $r_{\rm dark}$ is the detector dark current in units of photons per second.

First, we consider the case where Earth is unresolved from the Sun, detecting the laser as an anomalous spectral feature. In this case, DSOC is detectable against the full solar background at 150 AU. However, at such a nearby distance, Earth is resolvable from the Sun. To determine solar background in a semi-resolved case, we adopt a Moffat profile PSF model with a FWHM of 0.4 arcseconds (for good seeing) and power index $\alpha$ of 5. In this case, our $d_{\rm detect, laser}$ becomes 1.8 parsecs.

The ideal approach for detecting lasers like ours may be to use coronography to block light from the Sun. However, current coronagraph technologies do not provide high enough spectral resolution for this to be a feasible approach for narrow-band laser detection. Proposed and planned instruments like the High-resolution Infrared Spectrograph for Exoplanet Characterization (HISPEC\footnote{\url{https://www.ucobservatories.org/projects_at/keck-hispec/}}) may achieve this in the near future, further increasing the detectable distance for optical lasers.

\subsection{Heat Islands}
\label{ssec:heat_islands}

Most major current-day human cities are hotter than their rural or natural surroundings. This ``urban heat island'' effect makes them more emissive in the infrared, and can be quantified by the \ac{UHI}, essentially how many degrees hotter a city is than its surrounding environment \citep{piracha2022urban}. The city with the highest \ac{UHI} is Hong Kong, but estimates of its \ac{UHI} vary. \cite{zheng2023investigating} report up to 8$^{\circ}$~C in unfavorable weather conditions, while \citet{yee2022drivers} survey remote sensing data to find a 2.8$^{\circ}$~C maximum. We opt to use the ``best-case scenario'' of the most extreme published estimate: 10.5$^{\circ}$ C \citep{memon2009investigation}. 

Detecting these heat islands at a distance is challenging. One potential methodology is detailed by \citet{kuhn2015global}, where the disk-integrated thermal flux of over time can be analyzed by applying a 2D inversion to the lightcurve of a rotating planet, the same technique used in, e.g., sunspot detection in optical stellar lightcurves \citep[e.g., ][]{jarvinen2018mapping}. However, \citet{kuhn2015global} find that the natural variation in 10~$\upmu m$ flux for an Earth-like planet, even seen edge-on and correcting for the reflected light from the Sun, is 50X higher than the variation contributed by all of the integrated UHIs from current-day Earth. Therefore, it is not feasible to detect Earth-level heat islands in unresolved infrared time series.

However, if the disk of a planet is \textit{spatially-resolved}, then ``hot pixels'' from heat islands could potentially be measurable against the thermal background. Based on the discussion above, we use Hong Kong as our example \ac{UHI}, with a $\Delta T_{UHI}$ = $10.5^{\circ}$~C and a city area of $A = 1104$~km$^2$. Hong Kong has an average yearly temperature of $T_{city}$ = 296~K, which has a blackbody emission peak of $\sim10\upmu$m. The most sensitive detector for spatially-resolved emission in this part of the near-infrared is JWST's MIRI instrument, with its F1000W filter which covers $\sim$9--11$\upmu$m.

To compare flux from the city to that of an adjacent non-city area, we adopt $T_{noncity} = T_{city} - \Delta T_{UHI}$. In the 9--11$\upmu$m range, the city would produce almost 20\% more flux than the neighboring area. 
MIRI has a sensitivity\footnote{\url{https://jwst-docs.stsci.edu/jwst-mid-infrared-instrument/miri-performance/miri-sensitivity}} of 5 $\times$ $10^{-4}$~mJy for a required \ac{SNR} of 5 and an observation length of $\tau_{obs}$ = 10,000~sec. It also has a pixel scale of 0.11~arcsec/pixel, so the city would fill approximately a single pixel when observed from a distance of 0.416~AU. Beyond this distance, the city's excess flux in the pixel will be diluted by neighboring, non-UHI pixels. For small temperature changes $\Delta T$, $\Delta T \propto \Delta F$. Thus, for any given linear-dilution factor $f$, the temperature of the pixel containing the city will become:

\begin{equation}
\label{eq:dilution}
    T_{city, diluted} = \frac{(f^2 - 1)T_{noncity} + T_{city}}{f^2}
\end{equation}

Taking the diluted flux into account, we find that JWST MIRI could differentiate the heat island from Hong Kong from a nearby pixel up to a linear dilution factor of 72, corresponding to a distance of $d_{\rm detect, UHI}$ = $\sim 30$~AU. We note that this wavelength range lies in the N-band atmospheric window, where atmospheric transmission averages around $\sim$80\%, implying that the signal would be additionally diluted.

\section{Objects in Space or on Other Planetary Surfaces}
\label{sec:objects}

\subsection{Interstellar/Interplanetary Probes}
\label{ssec:radar}

Humans have left technosignatures in the solar system in the form of satellites that are in Earth's orbit and spacecraft on, and between, the other bodies of the solar system. Many of these objects are most detectable by their electromagnetic emissions (e.g., radio downlinks), but non-transmitting objects may only be detectable through radar. Detecting object-in-space technosignatures via radar is difficult, however, because received radar power scales inversely with the fourth power of the distance. There are three factors to consider when trying to maximize the radar-detectable distance of a technosignature: 1) the strength of the original radar transmission, 2) the reflectivity and surface area of the target, and 3) the sensitivity of the receiver.

As mentioned in Section \ref{sec:radio_transmissions}, the 2.4~GHz planetary radar system on the 305~m Arecibo Observatory was the strongest radio transmitter that has ever been built. The most radar-detectable objects that humanity has ever sent into space were the two satellites launched as part of the Echo project in the 1960s: Echo 1, launched in 1960 and Echo 2, launched in 1964 \citep{elder1995out}. Both satellites had the same general form-factor: reflective balloons made of mylar-polyester film, inflated to diameters of 30.48~m (Echo 1) and 41.1~m (Echo 2). These two missions had multiple goals: to demonstrate the potential of passive communication reflectors e.g. in two-way communication, to measure atmospheric density and solar pressure via their orbital changes \citep[given their high surface area and relatively low mass; ][]{zadunaisky1961experimental}, and to use geodesy to precisely measure Earth's shape. These mission goals dictated the huge surface area and reflectivity of the spacecraft, which make them ideal for radar detection. We will use the larger diameter of Echo 2 in the following calculations.

For objects at short distances, a \textit{monostatic} strategy is often applied, where the radar transmitter and the radio receiver are part of the same instrument --- this was commonly done with Arecibo \citep[e.g., ][]{campbell2022arecibo}. However, as we are trying to maximize the distance between the transmitter and the target, it should be noted that a \textit{bistatic} strategy might be required; Earth will rotate as the signal travels, and the travel time could be significant enough to require a different receiver at a another position on the planet. One modern bistatic setup uses Arecibo/Goldstone as the transmitter and the 100~m \ac{GBT} as the receiver \citep[e.g., ][]{rodriguez2021improved}. We will assume that the equivalent of a monostatic Arecibo-to-Arecibo setup can be used in this scenario, as the ``ideal'' standard of radar astronomy.

Here, we make a number of assumptions, most of which are relatively standard in radar astronomy. We assume that either the object's location is known or the area of the sky is being surveyed such that the radar beam is perfectly aligned with the center of the object, to maximize the final detection \ac{SNR}. 

The Earth-detecting-Earth calculation for radar is a bit lengthy and is described in full in Appendix \ref{sec:appendix}. We can fill in Equation \ref{eq:radar_distance_simple} with Equations \ref{eq:noise_power}--\ref{eq:radar_cross_section} to calculate the detectable radar range, taking into account both the transmitted power hitting the object and the power and detectability of the object's echo. We find that $d_{\rm detect, radar}$ = 0.145~AU for the Arecibo-to-Arecibo monostatic scenario for 1~hour of integration time.

\subsection{Objects on Planetary Surfaces}
\label{ssec:rover_imagery}

Technosignatures on planetary surfaces will be easiest to identify if they are large and have sufficiently distinct characteristics from the planetary background. \citet{osmanov2024we} recently considered this problem for \acp{ETI} which have observational capabilities (i.e., optical interferometers) that far exceed current astronomical technologies; here, we focus instead on $\iota$ = 1 observing technologies. The largest humanmade objects on a non-Earth surface are the Apollo landing modules, which are about 4~m in scale \citep{orloff2000apollo}, followed by the $\sim$3~m scale rovers such as Curiosity \citep{grotzinger2012mars}. Non-terrestrial technosignatures like these are made primarily of metal, leading to potentially large differences in reflectivity between the objects and the surfaces they sit upon. For example, Apollo 11 landed on a lunar mare, one of the darker, low-lying plains on the Moon's surface. Numerically simulating the albedo distribution of the lunar maria from \citet{saari1972sunlit} in $10^6$ trials provides an average albedo of $\alpha_{maria}$ = 0.068 for the maria and $\alpha_{uplands}$ = 0.1216 for the uplands. We will assume the optimistic scenario of an aluminum artifact \citep[$\alpha_{Al} = $0.85, ][]{ruhland2019effects} against a fully solar-illuminated mare.

The best existing tool to search for this kind of signature is a high-resolution imager; the \ac{LRO} \ac{NAC} provides images with the finest spatial-resolution (down to $\sim$0.5~m) of any non-Earth planetary surface \citep{robinson2010lunar}. \ac{LRO} orbits at 50~km from the moon's surface, and is designed to detect the lunar surface with \ac{SNR}$_{moon}$ = 50--200, implying a brightness accuracy of $\Delta S = 0.005$ at best. Assuming an Earth-Sun distance of 1~AU and an Earth-Moon distance of 3.84 $\times$ 10$^5$~km, we can estimate a best-case solar flux of 1379~W/m$^2$ falling upon the lunar maria (assuming a solar power of 3.86 $\times 10^{26}$W). We can use the albedo difference between the maria and the artifact, as well as the same pixel-dilution formulation from Equation \ref{eq:dilution}, to estimate the distance at which the \ac{SNR} difference between neighboring pixels is 5 (i.e., 5$\Delta S$). We find that this happens at a dilution factor of 21.5, leading to a maximum detectable distance for planetary surface artifacts of $d_{\rm detect, surface} = $8600~km with current technology.

\subsection{Clarke Belts: Satellites Observed via Transits}
\label{ssec:satellites}

Large collections of artificial satellites orbiting a planet may be visible as a diffuse cloud with a geostationary ring-like structure \citep[i.e., a ``Clarke exo-belt'' as defined in][]{socas2018possible} which would affect the planet's transit signature as it passes in front of a star. Modern transit instruments are not built for the level of saturation that this strategy would require \citep[even at interstellar distances, see][]{sallmen2019improved}, and with the extremely low density of satellites in Earth orbit compared to extrapolations in previous work (lower by a factor of $\sim$10$^6$--10$^7$), Earth's satellite belt is an ``undetectable'' signature in the Earth-detecting-Earth framework. 

However, for a relatively nearby observer, an individual satellite may be detected by radar reflection (as discussed in Section \ref{ssec:radar}) or through transit imaging. In a similar fashion to the pixel dilution approaches of Sections \ref{ssec:heat_islands} and \ref{ssec:rover_imagery}, the signature of a satellite in transit across the Sun may be detectable if enough light is blocked for an SNR=5 dip in a pixel's brightness. For this calculation, we adopt the International Space Station (ISS) as our satellite, assuming a surface area of 3,000 square meters (based on the 2,500 square meter size of the solar array\footnote{\href{https://www.nasa.gov/image-article/solar-arrays-international-space-station-2/}{https://www.nasa.gov/image-article/solar-arrays-international-space-station-2/}}). 

To avoid the problem of saturation that most state-of-the-art telescopes would be subject to if pointed at the sun, we select a solar telescope: the Daniel K. Inouye Solar Telescope \citep[DKIST,][]{Rimmele2020}. For this calculation, we adopt DKIST's collecting area (a 4m diameter) and its Visible Broadband Imager's wavelength range (blue channel, 393.3–486.4 nm), quantum efficiency ($>$60\%), and pixel scale (0.011$\arcsec$) to determine the solar (approximated as a 5800 Kelvin blackbody) photon detection rate on a single pixel as a function of distance. DKIST's observations are typically stacked for totals of $\sim3.2$ seconds. Here, we adopt an optimal observation time as the time it takes Earth to transit a pixel, so that the fractional blockage of the pixel's light can be assumed constant throughout the observation. Thus, we arrive at a detectable distance $d_{\rm detect,transit}=1.3$~AU for an observation time of 0.4~seconds. Strategic image stacking to track the satellite spatially may make detection possible at larger distances, effectively increasing observation time without causing further signal dilution.

\section{Results}
\label{sec:results}

\begin{figure}
    \centering
    \includegraphics[width=\textwidth]{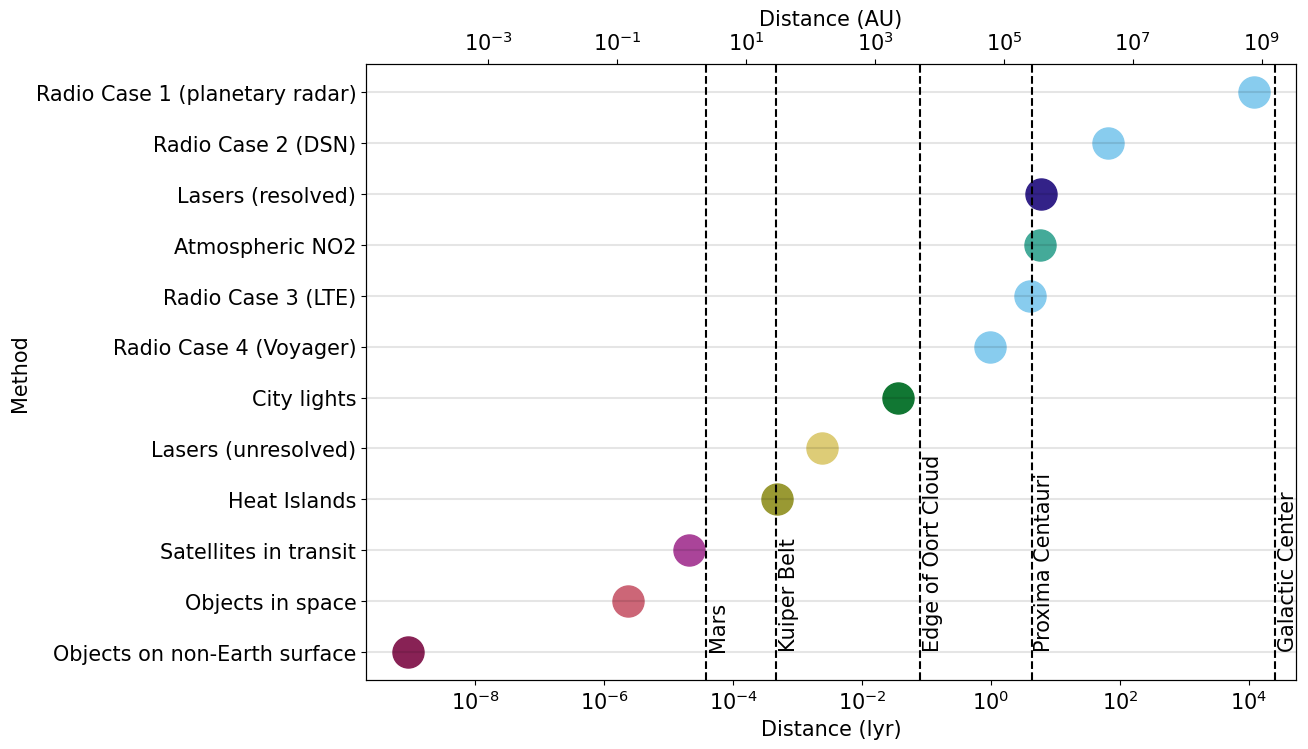}
    \caption{The maximum distances that each of Earth's modern-day technosignatures could be detected at using modern-day receiving technology, in visual form. Also marked are various astronomical objects of interest.}
    \label{fig:only_fig}
\end{figure}

Collating the results from Sections \ref{sec:radio_transmissions}--\ref{sec:objects}, we find a range of ``Earth detecting Earth'' detectability distances as listed in Table \ref{tab:toplevel_table} and illustrated in Figure \ref{fig:only_fig}. From these summaries, we can derive the following takeaways:

\textbf{Takeaway \#1: Radio remains the way that Earth is most detectable at $\iota$ = 1} 

This takeaway is not much of a surprise, because \ac{SETI} was first proposed and developed in the radio regime for exactly this reason \citep{Cocconi1959}. However, as a myriad of new technosignature search approaches have been popularized in recent years, it is a good reminder that the advantages of radio technosignature searches still hold over 60 years later, and that radio searches give us the best chance of detecting an Earth-level \ac{ETI} (by a factor of four \textit{orders-of-magnitude} in the most extreme cases).

\textbf{Takeaway \#2: Investment in atmospheric biosignature searches has opened up the door for atmospheric technosignature searches} 

Part of the reason that atmospheric NO$_2$ is detectable at such a large distance is that humans have greatly increased its concentration in our atmosphere. A more vital part is that incredible amounts of investment and technological development have gone into spectroscopic characterization of exoplanet atmospheres in the past two decades, in both transmission and absorption, a trend that shows no sign of stopping given the recommendation for a NASA flagship exoplanet characterization mission in the Astronomy and Astrophysics 2020 Decadal Survey \citep{decadal}, and NASA's subsequent embrace of the Habitable Worlds Observatory.  This point also highlights the overlap between biosignatures and technosignatures as scientific projects. The same wavelength bands for which HWO is optimized in its biosignature search strategy will also be those that include potential technosignatures.

\textbf{Takeaway \#3: Humanity's remotely-detectable impacts on Earth and the solar system span 12 orders-of-magnitude} 

This is a good reminder that useful technosignatures are characterized by being detectable with astronomical instrumentation, and function more on astronomical scales than human ones. While some of the technosignatures with shorter detection distances at $\iota$ = 1 have strong qualities in other parts of the Axes of Merit (see Section \ref{sec:discussion}), the challenge of detectability is paramount when restricting to ``Earth-level'' \ac{ETI}s.

\textbf{Takeaway \#4: Our modern-day planetary-scale impacts are modest compared to what is assumed in many technosignature papers}

For good reason, papers that theorize new technosignatures often imagine \ac{ETI}s with $\iota_{TS} \gg 1$ in making their case for the detectability of that signature. Many of them also rely on next-generation instrumentation that is still in the concept development phase \citep[e.g., the ExoLife Finder][]{berdyugina2018exo}, setting $\iota_{detector} \gg 1$ as well. With rising climate anxiety among the general public \citep[e.g., ][]{clayton2020climate}, it is useful to review what $\iota_{TS} = 1$ actually looks like for modern-day Earth; it is simultaneously true that humans have caused remotely-detectable global impacts, but also that not all of those impacts are catastrophically large or obvious from space. We should also be careful about extrapolating current technosignatures to scales of $\iota_{TS} = 10$ (or even $\iota_{TS} = 2$) without considering the changing context in which these technologies are being developed, used, and (sometimes) mitigated or phased-out \citep[e.g., the recovery of the ozone hole, ][]{kuttippurath2017signs}.  As another example, we are becoming aware of the negative health effects of \ac{UHI} \citep[as detailed in, e.g., ][]{piracha2022urban}; thus, work may be done to mitigate the concentrated regions of high infrared flux discussed in Section~\ref{ssec:heat_islands}. 

\textbf{Takeaway \#5: We have a multi-wavelength constellation of technosignatures, with more of the constellation becoming visible the closer you are} 

The technosignatures in Table~\ref{tab:toplevel_table} span radio, optical, and infrared wavelengths, and vary in size from a lunar landing module to the atmosphere of an entire planet. When confirming a biosignature or technosignature detection, these sorts of constellations, with independent lines of evidence and detection methods, will be critical in giving credence to the claim of the discovery of life. 

\begin{table}[ht]
    \centering
    \begin{tabular}{|c|c|c|c|}
    \hline
    Method  &   Distance (AU)   & Distance (lyr) &  Section  \\
    \hline
    
    Objects on non-Earth surfaces &  5.7 $\times$ $10^{-5}$ & 9.1 $\times$ $10^{-10}$ & \ref{ssec:rover_imagery} \\ 

    Objects in space & 1.4 $\times$ $10^{-1}$ & 2.3 $\times$ $10^{-6}$ & \ref{ssec:radar} \\

    Satellites in transit & 1.3 $\times$ $10^{0}$ & 2.1 $\times$ $10^{-5}$ & \ref{ssec:satellites} \\

    Heat islands & 3.0 $\times$ $10^1$ & 4.7 $\times$ $10^{-4}$ &  \ref{ssec:heat_islands} \\   

    Lasers (unresolved) & 1.5 $\times$ $10^2$ & 2.3 $\times$ $10^{-3}$ & \ref{ssec:lasers} \\

    City lights & 2.3 $\times$ $10^3$ & 3.6 $\times$ $10^{-2}$ & \ref{ssec:city_lights} \\  

    Radio Case 4 (Voyager) & 6.1 $\times$ $10^4$ & 9.7 $\times$ $10^{-1}$ & \ref{sec:radio_transmissions}  \\

    Radio Case 3 (LTE) & 2.5 $\times$ $10^5$ & 4.0 $\times$ $10^{0}$ & \ref{sec:radio_transmissions} \\
    
    Atmospheric NO$_2$ & 3.6 $\times$ $10^5$ & 5.7 $\times$ $10^{0}$ & \ref{sec:atmospheric} \\

    Lasers (resolved) & 3.7 $\times$ $10^5$ & 5.9 $\times$ $10^{0}$ & \ref{ssec:lasers}\\

    Radio Case 2 (DSN) & 4.1 $\times$ $10^6$ & 6.5 $\times$ $10^{1}$ & \ref{sec:radio_transmissions} \\
    
    Radio Case 1 (planetary radar) & 7.5 $\times$ $10^8$ & 1.2 $\times$ $10^{4}$ & \ref{sec:radio_transmissions} \\

    \hline
    \end{tabular}
    \caption{The maximum distances that each of Earth's modern-day technosignatures could be detected at using modern-day receiving technology, in increasing distance order.}
    \label{tab:toplevel_table}
\end{table}

\section{Discussion}
\label{sec:discussion}

\subsection{On the use of \texorpdfstring{$\iota$}{iota}=1}
This work focuses on $\iota$ = 1, which strictly limits the kinds of technosignatures we consider, their scale, and the detection technologies that we can use to find them. It is likely that, statistically, if we find an \ac{ETI}, then it will have been around for much longer than humans have been on Earth \citep{kipping2020contact}. Similarly, \ac{ETI}s that are the most detectable will likely be those that are employing large, obvious technologies which far exceed anything that we have built on Earth \citep[e.g., Dyson Spheres, ][]{wright2020dyson}. Our progress in building better telescopes and detectors is also non-negligible, especially over the potentially decades- or centuries-long timescale that may be necessary for the search for life in space.

So why bother focusing on $\iota$ = 1? First, from a historical perspective, the insight that kicked off modern \ac{SETI} was that humanity could, with great effort, communicate with \textit{ourselves} at interstellar distances \citep{Cocconi1959}. The $\iota$ = 1 framing \edit1{was sensible in the context of a newly developed technology, despite an implicit understanding that the  ``Earth 1960'' technology was a starting point from which radio technology would continue to grow and evolve. We are now approaching another moment of importance in radio technosignatures}, as described in Section~\ref{sec:radio_transmissions}, where we will soon be able to detect Earth-level radio leakage from nearby systems. This profound transition will open up a part of parameter space with radio signals that are not intentionally sent, but are persistent, omnidirectional, and most importantly, represent power levels that humanity has already demonstrated in our own technology. Second, this paper puts wildly disparate technosignatures on an equal footing using the ichnoscale framework so they can be compared quantitatively for the first time. Third, $\iota$ = 1 showcases a series of useful, theoretical outcomes. The strictness about the allowed capabilities grounds the discussion of which technosignature searches are likely to have potential for success. Additionally, the extension of the ``SETI as a mirror'' concept \citep{tarter2010seti} provides an updated picture of humanity's technosphere. Finally the focus on planetary-scale technosignatures provides very specific suggestions for which searches to pursue in a universe where large-scale energy expenditures and/or travel between systems is logistically infeasible. While science fiction is, for example, replete with mechanics for rapid interstellar travel, all current physics implies it would be slow and expensive. We should take that constraint seriously. 

Using the $\iota$ = 1 framework as the only paradigm, i.e., using these results to definitively argue for or against certain technosignatures, would be a misinterpretation of the work. Future work should consider similarly constrained but entirely orthogonal scenarios (e.g., maximize the axis of Detectability instead, focusing on galactic-scale technosignatures). It is also worth noting that we do not consider the difficulty of distinguishing a technosignature from abiotic and biotic confounders in this work, as this is a separate axis in the Nine Axes (``ambiguity'').

\subsection{Implications for Solar System Science}

Something that should stand out in Figure \ref{fig:only_fig} is that objects in space and objects on non-Earth surfaces are only detectable at distances smaller than the distance between Earth and Mars. This highlights the fact that planetary surfaces and interplanetary space are relatively underexplored. \Ac{LSST} will help fill in our understanding of free-floating objects in interplanetary space, opening new avenues for \ac{SETI} \citep{davenport2019seti, lacki2019shiny}, while the relative paucity of planetary surface imaging also provides an opportunity for new kinds of solar system \ac{SETI} which could themselves lead to the discovery of interesting geological anomalies \citep[e.g., ][]{davies2013searching}.

While generally radar is used to characterize nearby asteroids or terrestrial planetary surfaces, astronomical radar has actually been used for human technosignature detection before, at four lunar distances \citep{brozovic2017radar}. Arecibo and Goldstone were used to re-locate the \ac{SOHO} satellite and determine its spin rate after it lost attitude control and strayed from its Lagrange Point \citep{vandenbussche1999soho}. This provides a historical precedent for the use of radar to characterize free-floating technosignature objects.

\subsection{Additional Considerations for Atmospheric Technosignatures}
\label{ssec:atmosphere_discussion}

In Section \ref{sec:atmospheric}, we focus on NO$_2$ as a ``best-case'' $\iota$ = 1 atmospheric technosignature due to its abundance, its significant technological sources, and its features in a part of the electromagnetic spectrum that is accessible to an upcoming mission. However, the library of atmospheric technosignatures is much larger than just NO$_2$. \ac{CFCs} are a more unambiguous technosignature than NO$_2$, but, as with many technosignature gases, their abundances and spectral signatures will be more detectable around non-Sunlike hosts e.g., white dwarfs \citep[CFC-11, CFC-14][]{lin2014detecting} or M-dwarfs \citep[CFC-11, CFC-12][]{haqq2022cfc}. 

Constellations of atmospheric technosignatures could also indicate the presence of technological life. Much like the idea of ``biosignature pairs'' \citep{schwieterman2018exoplanet}, multiple atmospheric constituents which, on their own, would be interpreted as solely abiotic or biotic could \textit{together} be a technosignature. For example, NH$_3$ and N$_2$O together could serve as evidence of extraterrestrial agriculture \citep{Haqq-Misra2022agriculture}. 

Future work should calculate detectability for atmospheric technosignatures that have been proposed but not quantitatively characterized, e.g., CFC-113 \citep{schneider2010far} or the 558~nm O$_2$ airglow line from nuclear weapons detonation \citep{stevens2016observational}, as well as for the technosignature ``constellations'' mentioned above.

Atmospheric technosignature concentrations are not necessarily stable over time, even on the timescale of a human lifetime, adding both additional benefits and drawbacks to searching for these signatures. From a detectability standpoint, sometimes the \textit{change} in an atmospheric component over time can constitute its own technosignature, even if a static snapshot of the concentration in the atmosphere would not be a technosignature. For example, human technology is releasing CO$_2$ into the atmosphere at a rate that is 10$\times$ greater than the Paleocene-Eocene thermal maximum, one of the most rapid abiotic releases of CO$_2$ in Earth's recent geological history \citep{gingerich2019temporal}. While we do not have decades-long baselines for any current astronomical missions, this will change as biosignature and technosignature science continue. We can expect some exoplanets to be continuously targeted over the lifetime of multiple instruments and missions. This becomes important because, for example, over the last century not only has the total abundance of CO$_2$ increased but its rate of increase (currently 40~Gtons/yr) has itself been rapidly increasing. Thus the CO$_2$ time derivative is an $\iota$ = 1 technosignature for present-day Earth, and this remains a potential avenue for future dedicated atmospheric technosignature missions. However, there's a detectability drawback as well: as human industries like agriculture and manufacturing change their methods, sometimes in response to social pressure, some atmospheric technosignatures will diminish over time. The reversal of the non-equilibrium ozone depletion caused by human production of CFCs \citep{solomon1999stratospheric}, or even the decrease in NO$_2$ since 1980 discussed in Section \ref{sec:atmospheric}, are examples of short-lived atmospheric technosignatures that may continue to fall from detectable levels on the timescale of decades to centuries\footnote{Simultaneously, however, only 2 species of CFCs are banned and decreasing in concentration. The others keep going up because they're largely (chemically) inert \citep[e.g., ][]{western2023global}}.

\subsection{Other Technosignatures}

One key technosignature neglected thus far is the interstellar probe, as proposed by \citet{Bracewell1960}. At this time, a few interstellar probes containing information about humanity, including the Voyager and Pioneer probes, are on their way our of the Solar System. 
These objects, although also detectable via radar (see Section \ref{ssec:radar}), are not directly analogous to the technosignatures examined above because the Bracewell approach to interstellar communication essentially requires physical contact with the object. The Voyager probes may remain intact for billions of years \citep{Oberg2021}, but their reach is limited by their relatively small size, recent launch times, and much slower-than-light speeds.

We chose not to calculate a specific distance for surface geoengineering --- for example, mines, reservoirs, and other large, human-made structures upon Earth's surface for two reasons: first, they are equivalent in length scale to the heat islands discussed in Section \ref{ssec:heat_islands}, and second, they are more complex in their reflectance, visibility through the atmosphere, and distinctness from their environment than the heat island case. 

There are also plenty of technosignatures that are global but likely not remotely-detectable. The prevalence of microplastics in our oceans and soil is undeniable upon in-situ analysis \citep{andrady2011microplastics}, but not easily apparent at a distance\footnote{Microplastics can change the inherent optical properties and backscattering signal of ocean water's response to syntheic aperture radar \citep{salgado2021assessment}. This technique can be combined with ground-based measurements to study microplastics as a larger phenomenon for global mapping of marine litter \citep[e.g., ][]{davaasuren2018detecting, evans2021toward}}. Another example is radioactive isotopes in global stratigraphy from atmospheric nuclear tests in the 1960s \citep{waters2018global}.

\subsection{SETI as a Mirror}

Finally, in the spirit of using \ac{SETI} as a mirror, following the tradition of \citet[e.g., ][]{sullivan1978eavesdropping}, we will briefly discuss what information about Earth and humanity can be determined from each element of our technosignature constellation. Humans have intentionally sent objects, time capsules, and transmissions beyond our planet since it was technologically possible to do so. \citet{Quast2021} has considered what aggregate cultural message all of these disparate efforts (some serious \ac{ETI} communication attempts, some publicity stunts) would carry while \citet{gorman2005cultural} looks at the implications of our material ``spacescapes.'' However, we can broaden this question to go beyond intentional communication attempts and consider how we would be interpreted by a distant observer via the ```technomarker' activities that define our world at a distance'' from \citet{Quast2021}. It is possible for \acp{ETI} to hypothesize about our culture, society, biosphere, etc. from our unintentional technosignatures, and thinking through those possible hypotheses can help us interrogate how we are presenting ourselves to the galaxy: how we organize socially, how we relate to the world around us, how we perceive and experience things, and perhaps even what we value. It would be reductive to say that we might not be putting our best feet forward, as it is not possible for us to understand how \acp{ETI} would evaluate our technosignatures in any human moralistic framework, but this line of thinking does push us towards taking that question seriously and developing the skills necessary to notice when our anthropocentric biases enter into our analyses and hypotheses regarding \acp{ETI}.

We send many radio signals into space, over a range of powers, morphologies, frequencies, and duty cycles. While radio's importance to our planet is undeniable, a hypothetical observer could interpret this prevalence in a range of ways. They would likely make the unsurprising observation that our atmosphere is transparent to radio waves. They may infer that we cannot sense radio waves biologically, or the world would be in constant cacophony. Conversely, our reliance on radio waves could make it natural for an alien species to wonder if it's \textit{because} we can detect them biologically!  Strong pulses from planetary radar could be interpreted as communication with spacecraft beyond our atmosphere or (correctly) as radar to study other bodies in our solar system. Our atmospheric technosignatures, if monitored over time, also tell a fascinating tale: clearly, we're working on ``short'' timescales astronomically, where the increases in NO$_2$, CFCs, and other gases are indicative of acceleration and a period of rapid change. This could be interpreted as reckless industrialization  without an eye to the desirability of the future stable state. Alternately, the rapidity of the change could read as intentional coordinated terraforming. The depletion of the ozone layer, and then its subsequent recovery, also imply an amount of awareness about the impacts of our actions on our host planet. It would be straightforward to infer that we see in the visible wavelengths and desire nighttime illumination, based on our city lights. The presence of heat islands could be used to determine that we cluster into cities on our surface; if an observer were to see enough cities to infer a distribution, it could likely be cross-referenced with city light data to see a correlation with coastal areas. It is worth nothing that elements of our social organization and/or behavior could also be inferred from the detection of such artifacts.  The presence of islands of artificial lighting (i.e. cities) not only indicates that our biological sensors use visible light, it also demonstrates a particular mode of socially networked organization which itself can imply specific modes of hierarchy. Human-made objects in space, at the very least, imply that we know how to launch things into space, of a certain size scale. An external observer might be curious about the rapid rate satellites were appearing in Earth orbit (perhaps indicating more short-timescale thinking or, alternately, an intentionally grand orbital plan), and satellites in a geostationary orbit would imply a 24 hour rotation period. 

To summarize this exercise, Earth's present-day technosignatures hold clues to humanity's culture, society, and biosphere. Some of these clues may be straightforward to interpret correctly if the associated technosignature is detected, but others could be interpreted in wildly varying ways, as there are many activities or prior states that could produce the same astronomically observable outputs. We should keep this in mind as we ourselves hypothesize about the ETIs behind any future technosignature candidates.

\section{Conclusion}
\label{sec:conclusion}

In this paper, we investigated the distances at which technosignatures of the modern-day Earth could be detected with the astronomical instrumentation of the modern-day Earth, in a formulation we called ``Earth Detecting Earth.'' This paradigm focuses on limiting extrapolation to unknown technologies to understand the detectability of Earth by its constellation of technosignatures. We put various technosignatures across the electromagnetic spectrum onto a quantified ``detectability'' distance scale, including radio transmissions from different scales of radio transmitters, optical emission from nightside city lights, focusing lasers for adaptive optics, heat islands from cities, free-floating objects that could be pinged with radar, the belt of satellites around the planet, and objects such as rovers on non-Earth planetary surfaces. 

The results of this investigation are shown in Figure \ref{fig:only_fig}, but in summary:

\begin{itemize}
    \item Radio remains the way that Earth is most detectable at $\iota$ = 1
    \item Investment in atmospheric biosignature searches has opened up the door for atmospheric technosignature searches
    \item Humanity's remotely-detectable impacts on Earth and the solar system span 12 orders-of-magnitude
    \item Our modern-day planetary-scale impacts are modest compared to what's assumed in many technosignature papers
    \item We have a multi-wavelength constellation of technosignatures, with more of the constellation becoming visible the closer the observer becomes
\end{itemize}

This should not be the only way of characterizing and sorting a portfolio of technosignature approaches, but will be complemented by future work emphasizing aspects perpendicular to the minimal-extrapolation approach taken here.

\acknowledgments

The authors thank Michael Busch for his assistance and insight in the radar astronomy portions of this work, and Jessica Lu for her insight on the laser portion. S.Z.S, M.J.H., P.F., J.T.W., and R.K. gratefully acknowledge support from the NASA Exobiology program under grant 80NSSC20K0622. S.Z.S. acknowledges that this material is based upon work supported by the National Science Foundation MPS-Ascend Postdoctoral Research Fellowship under Grant No. 2138147. The Penn State Extraterrestrial Intelligence Center and the Center for Exoplanets and Habitable Worlds are supported by the Pennsylvania State University and the Eberly College of Science. Computations for this research were performed on Penn State’s Institute for Computational and Data Sciences’ Roar supercomputer. M.J.H.'s work was partially supported by the Pennsylvania Space Grant Consortium Graduate Fellowship.

\bibliography{references.bib}

\appendix
\section{Radar Formulation}
\label{sec:appendix}

We will start with a standard form of the radar range equation, as contained in \citet[e.g., ][]{ostro1993planetary}. The \ac{SNR} of a radar observation, in its most general form is:

\begin{equation}
    \mathrm{SNR} = \frac{P_{r}}{\Delta P_N}
\end{equation}

Where $P_r$ is the received power after the radar reflection and $\Delta P_N$ is the magnitude of the standard deviation of the random fluctuations in the noise level $P_N$ (which can just be measured and removed as a constant offset). The received power term consists of parameters related to the transmitter, the target, and the receiver:

\begin{equation}
\label{eq:radar_received_power}
    P_r = \frac{P_t G_t G_r \lambda^2 \sigma}{(4\pi)^3 R^4}
\end{equation}

Where $P_t$ was the original transmitted radar power, $G_t$ and $G_r$ are the gains of the transmitter and receiver, respectively ($G_t$ = $G_r$ for a monostatic radar), $\lambda$ is the wavelength of the radar transmission, $\sigma$ is the radar cross-section of the target, and $R$ is the distance from the transmitter to the target. Meanwhile, $\Delta P_N$ can be expanded, in the assumption of Gaussian noise, as:

\begin{equation}
    \label{eq:radar_noise_fluctuations}
    \Delta P_N = \frac{P_N}{\sqrt{\Delta f \Delta t}} 
\end{equation}

Where $P_N$ is the noise power, which will determine the scale of the fluctuations, $\Delta f$ is the frequency resolution of the data, and $\Delta t$ is the integration time of the data.

In this project, we are of course interested in deriving a maximum distance, so we will rearrange Equation \ref{eq:radar_received_power} and combine it with Equation \ref{eq:radar_noise_fluctuations} to get the following:

\begin{equation}
   \label{eq:radar_distance_simple}
    R = \bigg(\frac{P_t G_t G_r \lambda^2 \sigma \sqrt{\Delta f \Delta t}}{(4\pi)^3 (\mathrm{SNR}) P_N} \bigg)^{\frac{1}{4}}
\end{equation}

The noise power $P_N$ can be written as:

\begin{equation}
\label{eq:noise_power}
    P_N = kT_{sys}\Delta f
\end{equation}

Where $k$ is the Boltzmann constant and $T_{sys}$ is the system temperature of the receiver. In the ideal case, the frequency resolution of the data would be perfectly matched to the frequency spread of the return echo and no larger (to minimize the noise contribution). In this case, we can write:

\begin{equation}
    \Delta f = \frac{D}{\lambda P_{rot}}
\end{equation}

Where D is the diameter of the target object and $P_{rot}$ is the rotation period of the target (which can produce a Doppler smearing of a radar pulse that is largest at the edges of the target).

We can also return to the gain terms $G_r$ and $G_t$, which can be expanded into instrument parameters as:

\begin{equation}
   G = \frac{4\pi \epsilon A}{\lambda^2}
\end{equation}

Where $\epsilon$ is the aperture efficiency of our system with collecting area $A$ at wavelength $\lambda$.

Finally, we can address the radar cross-section of the target, which is a straightforward combination of its diameter D and albedo $\alpha$ for a spherical object:

\begin{equation}
    \label{eq:radar_cross_section}
    \sigma = \frac{\alpha \pi D^2}{4}
\end{equation}

\begin{acronym}
\acro{SETI}{the Search for Extraterrestrial Intelligence}
\acro{SNR}{signal-to-noise ratio}
\acro{RFI}{Radio Frequency Interference}
\acro{BL}{Breakthrough Listen}
\acro{ProxCen}{Proxima Centauri}
\acro{UWL}{Ultra-Wide  bandwidth,  Low-frequency receiver}
\acro{NEO}{Near-Earth Object}
\acro{ETI}{Extraterrestrial Intelligence}
\acro{ISS}{International Space Station}
\acro{DSN}{Deep Space Network}
\acro{GBT}{Green Bank Telescope}
\acro{METI}{Messaging Extraterrestrial Intelligence}
\acro{EIRP}{Effective Isotropic Radiated Power}
\acro{SKA1-Mid}{Square Kilometer Array Mid-Frequency}
\acro{SEFD}{System Equivalent Flux Density}
\acro{LOFAR}{Low Frequency Array}
\acro{CFCs}{chlorofluorocarbons}
\acro{UHI}{urban heat index}
\acro{LRO}{Lunar Reconnaissance Orbiter}
\acro{NAC}{Near-Angle Camera}
\acro{FAST}{the Five-hundred Meter Aperture Telescope}
\acro{LSST}{the Legacy Survey of Space and Time}
\acro{SOHO}{Solar and Heliospheric Observatory}
\end{acronym}

\end{document}